\documentclass[twocolumn,prl   ,superscriptaddress,10pt,aps,floatfix]{revtex4-2}
\usepackage[english]{babel}
\usepackage{graphicx}
\usepackage{amsmath}
\usepackage{amssymb}
\usepackage[colorlinks,linkcolor=blue,citecolor=blue,urlcolor=blue]{hyperref}
\usepackage{color}
\usepackage{soul}
\usepackage[normalem]{ulem}
\usepackage{xcolor}
\usepackage{makecell}
\usepackage{multirow}
\usepackage{orcidlink}
\definecolor{dblue} {RGB}{28,130,185}
\let\oldaddcontentsline\addcontentsline
\newcommand{\stoptocentries}{\renewcommand{\addcontentsline}[3]{}}
\newcommand{\starttocentries}{\let\addcontentsline\oldaddcontentsline}
\definecolor{nred}{RGB}{224,0,0}
\definecolor{nblue}{RGB}{28,130,185}
\definecolor{darkgreen}{rgb}{0,0.60,.2}
\definecolor{pgreen}{rgb}{0,1.0,0.}

\newcommand\redsout{\bgroup\markoverwith{\textcolor{brown}{\rule[0.6ex]{4pt}{1.0pt}}}\ULon}

\setcellgapes{2pt}
\makegapedcells

\newcommand{\be}{\begin{equation}}
\newcommand{\ee}{\end{equation}}
\newcommand{\bs}{\begin{equation}\begin{aligned}}
\newcommand{\es}{\end{aligned}\end{equation}}
\newcommand{\bg}{\begin{gathered}}
\newcommand{\eg}{\end{gathered}}

\begin{document}
\title{Geometric pairing of Nambu--Goldstone modes in spacetime symmetry breaking}

\author{Aleksander G\l{}\'odkowski \orcidlink{0000-0003-3893-9112}}

\affiliation{Institute of Theoretical Physics, Faculty of Fundamental Problems of Technology, Wroc\l{}aw University of Science and Technology, 50-370 Wroc\l{}aw, Poland}

\begin{abstract}
We identify a geometric pairing mechanism in systems with spontaneously broken spacetime symmetries, whereby pairs of Goldstone fields become canonically conjugate and hybridize into a single type-B Nambu--Goldstone mode (NGM). The mechanism originates from deformations of the local spacetime volume element induced by Goldstone fluctuations, which endow the Goldstone manifold with a Berry curvature. Integrating this pairing with existing reductions, we propose a general counting formula for the number of gapless NGMs in many-body systems. We demonstrate the mechanism in a microscopic model
of a two-component Bose--Einstein condensate, where the dilaton and the
$U(1)$ Goldstone field combine into a single NGM with quadratic
dispersion.
\end{abstract}

\maketitle
\stoptocentries

\textit{Introduction.}---Universal properties of a many-body system can be inferred from its pattern of spontaneous symmetry breaking (SSB), without detailed knowledge of the underlying microscopic dynamics. In particular, it determines the spectrum of Nambu--Goldstone modes (NGMs), gapless collective excitations that govern the low-energy physics. This connection is made concrete by the celebrated Goldstone theorem, which establishes a correspondence between spontaneously broken continuous symmetries and NGMs \cite{Goldstone:1961eq,PhysRev.127.965}. For a relativistic vacuum that breaks \textit{uniform symmetries}---whose generators commute with spacetime translations---the number
of NGMs is exactly equal to the number of broken generators,
$N_{\rm NGM}=N_{\rm BG}$. However, this simple relation breaks down in nonrelativistic systems
\cite{Nielsen:1975hm}, or when some of the broken symmetries are
\textit{nonuniform}, meaning that their generators commute
nontrivially with spacetime translations \cite{Ivanov:1975zq}. For uniform symmetries, this problem was largely resolved by Brauner, Watanabe, and Murayama, who identified the general conditions under which pairs of Goldstone fields become canonically conjugate and combine into a single type-B NGM, leading to the modern type-A/type-B classification of NGMs \cite{Watanabe:2011ec,Watanabe:2012hr}. For nonuniform symmetries, the reduction of the NGM spectrum is usually understood in terms of inverse Higgs constraints (IHCs), which systematically eliminate inessential Goldstone fields \cite{Ivanov:1975zq,Low:2001bw,Watanabe:2013iia,Brauner:2014aha,GlodkowskiPhD}.

In this Letter, we show that spontaneously broken spacetime symmetries can generate type-B NGMs through a mechanism which we dub \textit{geometric pairing}. We identify the general criterion for the mechanism to occur and elucidate its geometric origin in terms of an emergent Berry curvature on the Goldstone manifold. Combining the mechanism with IHCs and the Brauner--Watanabe--Murayama (BWM) pairing, we propose a general formula for the number of gapless NGMs in many-body systems. The formula provides an efficient way to determine the NGM spectrum without resorting to microscopic calculations. Finally, we apply the counting formula to a microscopic model of a two-component Bose--Einstein condensate (BEC), where it predicts a single type-B NGM arising from the geometric hybridization of the dilaton and the $U(1)$ Goldstone field. This prediction is confirmed by an explicit derivation of the low-energy effective theory from the microscopic model.

\textit{Geometric pairing.}---To introduce the mechanism, we begin by isolating the sector of the symmetry algebra responsible for geometric pairing. We therefore consider mutually commuting spontaneously broken generators \(Q_a\) acting nontrivially on unbroken translations,
\be\label{eq:spacetime}
[P_\mu,Q_a]
=
(\sigma_a)_\mu{}^\nu P_\nu\,.
\ee
Applying the coset construction \cite{Coleman:1969sm,Callan:1969sn,Ivanov:1975zq}, we parameterize the coset space by the representative
\be
\Omega
=
e^{x^\mu P_\mu}e^{\pi^a Q_a}\,,
\ee
where \(\pi^a\) denote the Goldstone fields. Note that this choice of parametrization ensures that the Goldstone fields transform homogeneously under spacetime translations, $\delta_{P_\mu} \pi^a = -\partial_\mu \pi^a$. The Maurer--Cartan form has the general form
\be
\omega
\equiv
\Omega^{-1}d\Omega 
=
dx^\nu e_\nu{}^\mu
\left(
P_\mu
+
\nabla_\mu\pi^a Q_a
\right)\,,
\ee
where \(e_\nu{}^\mu\) is the coset vielbein and \(\nabla_\mu\pi^a\) the
covariant derivative. Using the algebra Eq.~\eqref{eq:spacetime} we identify 
\be
e_\mu{}^\nu
=
\left(e^{\pi^a\sigma_a}\right)_\mu{}^\nu\,,
\qquad
\nabla_\mu \pi^a
=
\left(e^{-\pi^a \sigma_a}\right)_\mu{}^\nu\partial_\nu \pi^a \,.
\ee
With this we can define an invariant volume form 
\be
dV \equiv  |e|  \, dx^0 \wedge dx^1 \wedge \dots \wedge dx^d \,, 
\ee
where $|e|  = \exp{ \left[ \pi^a \textrm{tr} \left( \sigma_a \right) \right]} $.
We therefore see that Goldstone fluctuations deform the local spacetime volume element. The strictly invariant part of the effective action can be written as
\be
S
=
\int dV
\Big[ c_a^\mu \nabla_\mu  \pi^a - \frac12 g_{ab}^{\mu \nu}\nabla_\mu \pi^a \nabla_\nu \pi^b + \cdots \Big]  \,,
\ee
where  $c_a^\mu
\equiv
\langle j_a^\mu(x) \rangle $ is the ground state expectation value of the
Noether current associated with the broken generator $Q_a$. 
We omit a constant term proportional to $dV$, since it would generally generate a tadpole spoiling stationarity of the $\pi^a=0$ point. We also neglect possible Wess--Zumino contributions \footnote{For the commuting broken generators, such terms cannot contribute to the Berry curvature.}.
Expanding around the origin,
\bs 
\nabla_\mu\pi^a
&=
\partial_\mu\pi^a
-
\pi^b(\sigma_b)_\mu{}^\nu\partial_\nu\pi^a
+ \cdots , \\
|e|
&=
1+\pi^b(\sigma_b)_\rho{}^\rho+\cdots ,
\es
one finds a geometric Berry term,
\be \label{eq:Berry}
S_{\mathrm{Berry}}
=
\frac12
\int d^{d+1}x\,
\Omega_{ab}\pi^a\dot\pi^b \,,
\ee
where
\bs \label{eq:2form}
\Omega_{ab}
&=
V_{a\mu} \langle j_b^\mu\rangle 
-
V_{b\mu}  \langle j_a^\mu\rangle \,, \\
V_{a\mu}
&=
(\sigma_a)_\rho{}^\rho\delta_\mu{}^0
-
(\sigma_a)_\mu{}^0 \,.
\es
The antisymmetric tensor $\Omega_{ab}$ defines a Berry curvature two-form,
\be
\Omega
= 
\frac12
\Omega_{ab}
d\pi^a\wedge d\pi^b \,,
\ee
which endows the Goldstone manifold with a (partially) symplectic structure.
Since the Berry curvature is generally degenerate, it induces a decomposition of the tangent space $\mathcal T$ into degenerate and symplectic directions. To make this precise, define $K \equiv \ker\Omega\,,$ together with a complementary subspace \(\bar K\) such that 
\be
\mathcal T
=
K
\oplus
\bar K \,. 
\ee
Then, the restriction \(\Omega|_{\bar K}\) is nondegenerate, so that \(\bar K\) forms a symplectic subspace. Goldstone fluctuations spanning \(\bar K\) therefore organize into canonically conjugate pairs, each giving rise to a single type-B NGM, whereas those spanning \(K=\ker\Omega\) remain unpaired and correspond to type-A modes.
In particular, the numbers of type-A and type-B NGMs are
\be \label{eq:type}
N_A
=
N_{\rm BG}
-
\operatorname{rank}\Omega \,,
\quad
N_B
=
\frac12
\operatorname{rank}\Omega \,,
\ee
where \(N_{\rm BG}\) denotes the number of broken generators. Hence, the total number of NGMs is
\be
N_{\rm NGM}
=
N_A+N_B
=
N_{\rm BG}
-\frac12\operatorname{rank}\Omega \,.
\ee
Each pairing therefore reduces the number of independent NGMs by one. Moreover, the rank of the Berry curvature Eq.~\eqref{eq:2form} is constrained by the spacetime dimension. Introducing the one-forms
\be \label{eq:one-forms}
\mathcal J^\mu
\equiv
\langle j_a^\mu\rangle d\pi^a\,,
\quad
\mathcal V_\mu
\equiv
V_{a\mu} d\pi^a\,,
\ee
the Berry curvature Eq.~\eqref{eq:2form} takes the compact form
\be
\Omega
=
\mathcal V_\mu \wedge \mathcal J^\mu \,.
\ee
Since $\Omega$ is constructed from the one-forms
$\{\mathcal V_\mu,\mathcal J^\mu\}$, its rank satisfies
\be
\operatorname{rank}\Omega
\le
2(d+1) \,.
\ee
The maximal rank is attained when these one-forms are linearly independent spanning a
$2(d+1)$-dimensional subspace.

The physical intuition behind the pairing mechanism becomes especially transparent whenever the spatial current expectation values vanish, so that $\langle j_a^\mu\rangle=\langle j_a^0\rangle \delta^\mu_0$. Note that this condition is necessarily satisfied in an isotropic vacuum whenever the broken generators transform as scalars under spatial rotations. In such cases, the Berry curvature two-form Eq.~\eqref{eq:2form} simplifies to
\be\label{eq:berryIsotropic}
\Omega_{ab}
=
\tilde \sigma_a \langle j_b^0\rangle 
-
\tilde \sigma_b \langle j_a^0\rangle \,,
\ee
where $\tilde \sigma_a \equiv (\sigma_a)_i{}^i$ denotes the spatial trace. When $\tilde\sigma_a\neq0$, fluctuations of the associated Goldstone field locally deform the spatial volume element, $|e|  = \exp{ \left[ \pi^a \tilde \sigma_a \right]}$. In the presence of a uniform background charge density, these deformations modify the charge enclosed within a local spatial region. In turn, a fluctuation in one Goldstone direction redistributes the background charge seen by subsequent fluctuations. Therefore, deformations along different Goldstone directions become noncommutative, and closed trajectories on the Goldstone manifold accumulate a geometric Berry phase, encoded in Eq.~\eqref{eq:Berry}. In terms of the one-forms Eq.~\eqref{eq:one-forms}, the Berry curvature Eq.~\eqref{eq:berryIsotropic} takes the decomposable form $\Omega = \mathcal V_0 \wedge \mathcal J^0 \,,$
so that the geometric pairing occurs whenever $V_0$ and $J^0$ are linearly independent.
The symplectic structure is then supported on a single two-dimensional subspace, implying 
\be 
\textrm{rank} \, \Omega \leq 2 \,.
\ee 
Therefore, when the spatial current expectation values vanish, geometric pairing can produce at most one type-B NGM.

Finally, in a Lorentz-invariant vacuum, one has $\langle j_a^\mu\rangle=0$ whenever the broken generators transform trivially under the Lorentz group and the geometric Berry curvature Eq.~\eqref{eq:2form} vanishes.

\textit{Counting gapless NGMs.}---In general, many-body systems exhibit a richer pattern of symmetry breaking than that described by Eq.~\eqref{eq:spacetime}, involving more general nonuniform symmetries, as well as nonabelian symmetries. In such systems, the number of independent NGMs can be reduced not only by the geometric pairing introduced above, but also by IHCs and the BWM pairing. To develop a unified understanding of NGMs in many-body systems, it is therefore necessary to integrate geometric pairing with these existing reduction mechanisms and understand their mutual interplay when they occur simultaneously. More specifically, our goal is to establish a universal counting formula for the number of gapless NGMs from the algebraic structure of the broken symmetries and the expectation values of the associated Noether currents. 

With this goal in mind, we now consider a more general pattern of symmetry breaking in which geometric pairing coexists with the IHCs and BWM reductions. The algebraic structure relevant for the discussion may schematically be written as
\bs \label{eq:nonuniformAlgebra}
[P_\mu,Q_a]
&=
(\sigma_a)_\mu{}^\nu P_\nu
+
(\lambda_\mu)_a{}^b Q_b
\,, \\ 
[Q_a,Q_b]
&=
f_{ab}{}^C Q_C + \tau_{ab}{}^\mu P_\mu\,.
\es
The matrices $\sigma_a$ and $\lambda_\mu$ control the geometric pairing and the possible IHC reductions, respectively. The second commutator encodes the contributions to BWM pairing. Here, $f_{ab}{}^C$ are the structure constants, with $Q_C=(Q_c,Q_I)$ comprising all broken and unbroken charge generators, while $\tau_{ab}{}^\mu$ allows the commutator of two broken generators to contain a translation. In writing Eq.~\eqref{eq:nonuniformAlgebra}, we suppress terms proportional to unbroken generators in the first commutator, since they do not alter the general structure of the counting formula. We also assume the absence of additional identity-valued central extensions, which are not associated with independent charge generators (see, e.g., \cite{Kobayashi:2014eqa,Kobayashi:2014xua}). 

We begin by discussing the role of IHCs. For a systematic characterization of the resulting Goldstone reduction, we follow the kernel construction \footnote{See Ref.~\cite{GlodkowskiPhD} for details of the kernel construction. A closely related characterization was used in Ref.~\cite{Hayata:2013vfa} to elucidate the relation between broken symmetries and independent elastic variables.}. The argument can be summarized as follows. The algebra Eq.~\eqref{eq:nonuniformAlgebra}, implies the presence of a linear term in the covariant derivative,
\be
\nabla_\mu \pi^a
\simeq 
\partial_\mu \pi^a
-
(\lambda_\mu^T)^a{}_b \pi^b
+\cdots .
\ee
In turn, Goldstone fields not belonging to the common kernel of the matrices $\lambda_\mu^T$ enter the Maurer--Cartan form algebraically and can be eliminated by imposing IHCs. The Goldstone directions that survive the IHC reduction therefore span
\be
\ker\Lambda
\equiv
\bigcap_\mu \ker\lambda_\mu^T \,.
\ee
 The analysis of Goldstone pairings must therefore be restricted to $\ker\Lambda$, since fluctuations outside this subspace are eliminated by IHCs irrespective of their symplectic structure. Furthermore, the elements of $\ker\Lambda$ are in one-to-one correspondence with the independent broken generators whose associated Goldstone fields survive the IHC reduction \cite{GlodkowskiPhD}. In what follows, with a slight abuse of notation, we use the same symbol $\ker\Lambda$ for both spaces.

Within the reduced Goldstone manifold, both geometric pairing and the BWM mechanism contribute additively to the total Berry curvature. The latter generates the contribution \cite{Watanabe:2011ec,Watanabe:2012hr}
\be \label{eq:BWM}
\rho_{ab}
=
f_{ab}{}^C \langle j_C^0\rangle + \tau_{ab}{}^\mu \langle T^0{}_\mu \rangle \,, 
\ee
where $T^0{}_\mu$ denotes the energy-momentum density.
The total Berry curvature is therefore
\be
\mathcal F_{ab}
=
\Omega_{ab}
+
\rho_{ab}\,,
\ee
where $\Omega_{ab}$ is the geometric pairing two-form Eq.~\eqref{eq:2form}. Finally, restricting $\mathcal F$ to $\ker\Lambda$, we arrive at the counting formula
\be \label{eq:counting}
N_{\rm NGM}
=
\dim\ker\Lambda
-
\frac12
\operatorname{rank}
\mathcal F\big|_{\ker\Lambda}
 \,,
\ee
where $\mathcal F\big|_{\ker\Lambda}$ denotes the total Berry curvature computed on the charge space restricted to the subspace of independent broken generators. The first term describes the IHC elimination whereas the second term is responsible for the pairing of the Goldstone fields leading to type-A and type-B NGMs 
\bs
 N_A
&=
\dim\ker\Lambda
-
\operatorname{rank}\mathcal F|_{\ker  \Lambda}  \,, \\
N_B
&=
\frac12
\operatorname{rank}\mathcal F|_{\ker  \Lambda}  \,.
\es

Whenever the spatial current expectation values vanish, the restricted Berry curvature
$\mathcal F|_{\ker\Lambda}$ is completely determined by the vacuum charge densities and the symmetry algebra,
\be \nonumber
\left(\mathcal F|_{\ker\Lambda}\right)_{ab}
=
\tilde\sigma_a \langle j_b^0\rangle
-
\tilde\sigma_b \langle j_a^0\rangle
+
f_{ab}{}^C \langle j_C^0\rangle +
\tau_{ab}{}^\mu \langle T^0{}_\mu \rangle \,.
\ee
If all current expectation values vanish, one has $\mathcal F|_{\ker\Lambda}=0$, and the counting formula Eq.~\eqref{eq:counting} gives $N_{\rm NGM}
=
\dim\ker\Lambda$. 

\textit{Two-component BEC.}---We now illustrate the geometric pairing mechanism and apply the counting formula Eq.~\eqref{eq:counting} to a two-component BEC in $2+1$ dimensions. The system is described by two complex bosonic fields $\psi_1$ and $\psi_2$, which we collect into the doublet $\psi=(\psi_1,\psi_2)^T$. The microscopic Lagrangian takes the form
\be \label{eq:microscopic}
\mathcal L
=
i\psi^\dagger\partial_t\psi
-
\frac{1}{2m} \Vec{\nabla} \psi^\dag \cdot \Vec{\nabla} \psi - V(\psi, \psi^\dag)\,,
\ee
with the scale-invariant potential
\be \label{eq:potential}
V(\psi, \psi^\dag) = \lambda \left( \psi^\dag \sigma^y \psi \right)^2 +
\gamma \left( \psi^\dag \sigma^z \psi \right)^2 \,,
\ee
where $\lambda,\gamma>0$ and $\lambda\neq\gamma$.
The potential is minimized whenever 
\be \label{eq:minimum}
\psi^\dag \sigma^y \psi = 0\,, \quad \psi^\dag \sigma^z \psi = 0 \,.
\ee 
Restricting to the finite-density branch with $v > 0$, the most general ground state configuration satisfying the conditions Eq.~\eqref{eq:minimum} is
\be \label{eq:groundState}
\langle \psi \rangle  = v e^{i \theta} \begin{pmatrix}
    1 \\
    \pm 1
\end{pmatrix} \,,
\ee 
so that the vacuum manifold is 
\be 
\mathcal M_{\rm vac} \simeq \mathbb R_+ \times S^1 \times \mathbb Z_2 \,. 
\ee 
Within each $\mathbb Z_2$ sector, the connected component of $\mathcal M_{\rm vac}$ is parametrized by one $\mathbb R_+$ coordinate and one $U(1)$
phase. To describe low-energy fluctuations, we promote these coordinates
to slowly varying Goldstone fields. Choosing the reference vacuum with
fixed $v$ and $\theta=0$, we parametrize fluctuations along the
corresponding connected component of the vacuum manifold as
\be \label{eq:parameterization}
\psi(x)
=
e^{\sigma(x)} e^{i\varphi(x)}
\langle\psi\rangle \,,
\ee
where $\sigma(x)$ and $\varphi(x)$ are the Goldstone fields that parametrize fluctuations along the
$\mathbb R_+$ and $S^1$ directions,
respectively.
Substituting Eq.~\eqref{eq:parameterization} into the microscopic Lagrangian Eq.~\eqref{eq:microscopic}, after expanding to quadratic order around
$\sigma=\varphi=0$, we obtain the following effective field theory
\be \label{eq:eft}
\mathcal L_{\textrm{EFT}}
=
4v^2\varphi\dot\sigma
-
\frac{v^2}{m}
\left[
(\partial_i\sigma)^2
+
(\partial_i\varphi)^2
\right] \,.
\ee
The first term is a Berry term, implying that the Goldstone fields $\sigma$ and $\varphi$ form a canonically conjugate pair corresponding to a single type-B NGM with quadratic dispersion $\omega=\frac{k^2}{2m}$. 

We now show that the existence of this mode follows directly from the geometric pairing mechanism. To see this, we analyze the pattern of spontaneous symmetry breaking in the microscopic theory and compute the geometric Berry curvature. At tree level \footnote{Loop corrections may generate a quantum scale anomaly in two spatial dimensions thereby lifting the flat radial direction. }, the Lagrangian Eq.~\eqref{eq:microscopic} is invariant under the Schr\"odinger group \cite{Niederer:1972zz,PhysRevD.5.377}. The corresponding generators are
\bs \label{eq:generators}
H&=-\partial_t\,,
\quad
P_i=-\partial_i\,,
\quad
J =- \epsilon_{ij} x_i\partial_j\,, \\
N&=i\,,
\quad
K_i=-t\partial_i+imx_i\,, \\
D&=2t\partial_t+x^i\partial_i+1\,, \\
C&=- t^2\partial_t - tx^i\partial_i - t
+\frac{im}{2}x^2 \,.
\es
It is then straightforward to verify that the generators
$H$, $P_i$, and $J$ annihilate the vacuum configuration
Eq.~\eqref{eq:groundState}, whereas $N$, $K_i$, $D$, and $C$
act nontrivially and are therefore spontaneously broken. In particular, acting with the broken generators on the ground state Eq.~\eqref{eq:groundState}, we find
\be
\begin{aligned}
N\langle\psi\rangle
&=
i\langle\psi\rangle\,,
\\
D\langle\psi\rangle
&=
\langle\psi\rangle\,,
\\
K_i\langle\psi\rangle
&=
mx_i N\langle\psi\rangle\,,
\\
C\langle\psi\rangle
&=
-t D\langle\psi\rangle
+
\frac{m}{2}x^2 N\langle\psi\rangle\,.
\end{aligned}
\ee 
The first two relations reflect the fact that $N$ and $D$ generate the phase and radial
directions of the vacuum manifold, $N\langle\psi\rangle=\partial_\theta\langle\psi\rangle$ and $D\langle\psi\rangle=v\partial_v\langle\psi\rangle$. On the other hand, the last two relations indicate that the broken generators $K_i$ and $C$ do not generate additional independent directions. This is a microscopic realization of the inverse Higgs mechanism, in which the boost and conformal Goldstone fields do not correspond to independent directions on the vacuum manifold, but merely provide a redundant parametrization of it. 

In the language of the kernel characterization, the independent
Goldstone directions identified above correspond to the subspace of
independent broken generators
\be
\ker\Lambda
=
\operatorname{span}\{N,D\}\,.
\ee
We now analyze the symplectic structure on the reduced vacuum
manifold. Since $N$ and $D$ commute, the BWM contribution
Eq.~\eqref{eq:BWM} vanishes. However, dilatations act
nontrivially on translations, $[D,P_i]=-P_i$, and therefore generate a
nonvanishing geometric two-form
\be
\Omega_{ND}
=
-2\langle j_N^0\rangle
=
4v^2\,,
\ee
in precise agreement with the symplectic structure of the effective
theory Eq.~\eqref{eq:eft}. Thus, the restricted Berry curvature has rank
two, and the counting formula Eq.~\eqref{eq:counting} predicts a single
type-B NGM. Accordingly, the dilatation and phase
Goldstone fields are canonically paired and hybridize into a single
type-B NGM. Importantly, this conclusion follows entirely from symmetry---the breaking pattern, the algebra, and the vacuum charge density---without relying on microscopic details of the model.

\textit{Acknowledgements.}---We thank Tomáš Brauner and Paweł Matus for their useful comments and feedback on the manuscript. This work is supported in part by the Polish National
Science Centre (NCN) Sonata Bis grant 2019/34/E/ST3/00405.

\bibliographystyle{apsrev4-2}
\bibliography{bibliography}

\begin{thebibliography}{20}%
\makeatletter
\providecommand \@ifxundefined [1]{%
 \@ifx{#1\undefined}
}%
\providecommand \@ifnum [1]{%
 \ifnum #1\expandafter \@firstoftwo
 \else \expandafter \@secondoftwo
 \fi
}%
\providecommand \@ifx [1]{%
 \ifx #1\expandafter \@firstoftwo
 \else \expandafter \@secondoftwo
 \fi
}%
\providecommand \natexlab [1]{#1}%
\providecommand \enquote  [1]{``#1''}%
\providecommand \bibnamefont  [1]{#1}%
\providecommand \bibfnamefont [1]{#1}%
\providecommand \citenamefont [1]{#1}%
\providecommand \href@noop [0]{\@secondoftwo}%
\providecommand \href [0]{\begingroup \@sanitize@url \@href}%
\providecommand \@href[1]{\@@startlink{#1}\@@href}%
\providecommand \@@href[1]{\endgroup#1\@@endlink}%
\providecommand \@sanitize@url [0]{\catcode `\\12\catcode `\$12\catcode `\&12\catcode `\#12\catcode `\^12\catcode `\_12\catcode `\%12\relax}%
\providecommand \@@startlink[1]{}%
\providecommand \@@endlink[0]{}%
\providecommand \url  [0]{\begingroup\@sanitize@url \@url }%
\providecommand \@url [1]{\endgroup\@href {#1}{\urlprefix }}%
\providecommand \urlprefix  [0]{URL }%
\providecommand \Eprint [0]{\href }%
\providecommand \doibase [0]{https://doi.org/}%
\providecommand \selectlanguage [0]{\@gobble}%
\providecommand \bibinfo  [0]{\@secondoftwo}%
\providecommand \bibfield  [0]{\@secondoftwo}%
\providecommand \translation [1]{[#1]}%
\providecommand \BibitemOpen [0]{}%
\providecommand \bibitemStop [0]{}%
\providecommand \bibitemNoStop [0]{.\EOS\space}%
\providecommand \EOS [0]{\spacefactor3000\relax}%
\providecommand \BibitemShut  [1]{\csname bibitem#1\endcsname}%
\let\auto@bib@innerbib\@empty
\bibitem [{\citenamefont {Goldstone}(1961)}]{Goldstone:1961eq}%
  \BibitemOpen
  \bibfield  {author} {\bibinfo {author} {\bibfnamefont {J.}~\bibnamefont {Goldstone}},\ }\href {https://doi.org/10.1007/BF02812722} {\bibfield  {journal} {\bibinfo  {journal} {Nuovo Cim.}\ }\textbf {\bibinfo {volume} {19}},\ \bibinfo {pages} {154} (\bibinfo {year} {1961})}\BibitemShut {NoStop}%
\bibitem [{\citenamefont {Goldstone}\ \emph {et~al.}(1962)\citenamefont {Goldstone}, \citenamefont {Salam},\ and\ \citenamefont {Weinberg}}]{PhysRev.127.965}%
  \BibitemOpen
  \bibfield  {author} {\bibinfo {author} {\bibfnamefont {J.}~\bibnamefont {Goldstone}}, \bibinfo {author} {\bibfnamefont {A.}~\bibnamefont {Salam}},\ and\ \bibinfo {author} {\bibfnamefont {S.}~\bibnamefont {Weinberg}},\ }\href {https://doi.org/10.1103/PhysRev.127.965} {\bibfield  {journal} {\bibinfo  {journal} {Phys. Rev.}\ }\textbf {\bibinfo {volume} {127}},\ \bibinfo {pages} {965} (\bibinfo {year} {1962})}\BibitemShut {NoStop}%
\bibitem [{\citenamefont {Nielsen}\ and\ \citenamefont {Chadha}(1976)}]{Nielsen:1975hm}%
  \BibitemOpen
  \bibfield  {author} {\bibinfo {author} {\bibfnamefont {H.~B.}\ \bibnamefont {Nielsen}}\ and\ \bibinfo {author} {\bibfnamefont {S.}~\bibnamefont {Chadha}},\ }\href {https://doi.org/10.1016/0550-3213(76)90025-0} {\bibfield  {journal} {\bibinfo  {journal} {Nucl. Phys. B}\ }\textbf {\bibinfo {volume} {105}},\ \bibinfo {pages} {445} (\bibinfo {year} {1976})}\BibitemShut {NoStop}%
\bibitem [{\citenamefont {Ivanov}\ and\ \citenamefont {Ogievetsky}(1975)}]{Ivanov:1975zq}%
  \BibitemOpen
  \bibfield  {author} {\bibinfo {author} {\bibfnamefont {E.~A.}\ \bibnamefont {Ivanov}}\ and\ \bibinfo {author} {\bibfnamefont {V.~I.}\ \bibnamefont {Ogievetsky}},\ }\href {https://doi.org/10.1007/BF01028947} {\bibfield  {journal} {\bibinfo  {journal} {Teor. Mat. Fiz.}\ }\textbf {\bibinfo {volume} {25}},\ \bibinfo {pages} {164} (\bibinfo {year} {1975})}\BibitemShut {NoStop}%
\bibitem [{\citenamefont {Watanabe}\ and\ \citenamefont {Brauner}(2011)}]{Watanabe:2011ec}%
  \BibitemOpen
  \bibfield  {author} {\bibinfo {author} {\bibfnamefont {H.}~\bibnamefont {Watanabe}}\ and\ \bibinfo {author} {\bibfnamefont {T.}~\bibnamefont {Brauner}},\ }\href {https://doi.org/10.1103/PhysRevD.84.125013} {\bibfield  {journal} {\bibinfo  {journal} {Phys. Rev. D}\ }\textbf {\bibinfo {volume} {84}},\ \bibinfo {pages} {125013} (\bibinfo {year} {2011})}\BibitemShut {NoStop}%
\bibitem [{\citenamefont {Watanabe}\ and\ \citenamefont {Murayama}(2012)}]{Watanabe:2012hr}%
  \BibitemOpen
  \bibfield  {author} {\bibinfo {author} {\bibfnamefont {H.}~\bibnamefont {Watanabe}}\ and\ \bibinfo {author} {\bibfnamefont {H.}~\bibnamefont {Murayama}},\ }\href {https://doi.org/10.1103/PhysRevLett.108.251602} {\bibfield  {journal} {\bibinfo  {journal} {Phys. Rev. Lett.}\ }\textbf {\bibinfo {volume} {108}},\ \bibinfo {pages} {251602} (\bibinfo {year} {2012})}\BibitemShut {NoStop}%
\bibitem [{\citenamefont {Low}\ and\ \citenamefont {Manohar}(2002)}]{Low:2001bw}%
  \BibitemOpen
  \bibfield  {author} {\bibinfo {author} {\bibfnamefont {I.}~\bibnamefont {Low}}\ and\ \bibinfo {author} {\bibfnamefont {A.~V.}\ \bibnamefont {Manohar}},\ }\href {https://doi.org/10.1103/PhysRevLett.88.101602} {\bibfield  {journal} {\bibinfo  {journal} {Phys. Rev. Lett.}\ }\textbf {\bibinfo {volume} {88}},\ \bibinfo {pages} {101602} (\bibinfo {year} {2002})}\BibitemShut {NoStop}%
\bibitem [{\citenamefont {Watanabe}\ and\ \citenamefont {Murayama}(2013)}]{Watanabe:2013iia}%
  \BibitemOpen
  \bibfield  {author} {\bibinfo {author} {\bibfnamefont {H.}~\bibnamefont {Watanabe}}\ and\ \bibinfo {author} {\bibfnamefont {H.}~\bibnamefont {Murayama}},\ }\href {https://doi.org/10.1103/PhysRevLett.110.181601} {\bibfield  {journal} {\bibinfo  {journal} {Phys. Rev. Lett.}\ }\textbf {\bibinfo {volume} {110}},\ \bibinfo {pages} {181601} (\bibinfo {year} {2013})}\BibitemShut {NoStop}%
\bibitem [{\citenamefont {Brauner}\ and\ \citenamefont {Watanabe}(2014)}]{Brauner:2014aha}%
  \BibitemOpen
  \bibfield  {author} {\bibinfo {author} {\bibfnamefont {T.}~\bibnamefont {Brauner}}\ and\ \bibinfo {author} {\bibfnamefont {H.}~\bibnamefont {Watanabe}},\ }\href {https://doi.org/10.1103/PhysRevD.89.085004} {\bibfield  {journal} {\bibinfo  {journal} {Phys. Rev. D}\ }\textbf {\bibinfo {volume} {89}},\ \bibinfo {pages} {085004} (\bibinfo {year} {2014})}\BibitemShut {NoStop}%
\bibitem [{\citenamefont {G{\l}\'odkowski}(2026)}]{GlodkowskiPhD}%
  \BibitemOpen
  \bibfield  {author} {\bibinfo {author} {\bibfnamefont {A.}~\bibnamefont {G{\l}\'odkowski}},\ }\emph {\bibinfo {title} {Effective Theories for Many-Body Systems with Nonuniform Symmetries}},\ \href {https://bip.pwr.edu.pl/strona-glowna/postepowania-o-nadanie-stopnia-doktora---publikowane-od-2023-r/aleksander-glodkowski} {Ph.D. thesis} (\bibinfo {year} {2026})\BibitemShut {NoStop}%
\bibitem [{\citenamefont {Coleman}\ \emph {et~al.}(1969)\citenamefont {Coleman}, \citenamefont {Wess},\ and\ \citenamefont {Zumino}}]{Coleman:1969sm}%
  \BibitemOpen
  \bibfield  {author} {\bibinfo {author} {\bibfnamefont {S.~R.}\ \bibnamefont {Coleman}}, \bibinfo {author} {\bibfnamefont {J.}~\bibnamefont {Wess}},\ and\ \bibinfo {author} {\bibfnamefont {B.}~\bibnamefont {Zumino}},\ }\href {https://doi.org/10.1103/PhysRev.177.2239} {\bibfield  {journal} {\bibinfo  {journal} {Phys. Rev.}\ }\textbf {\bibinfo {volume} {177}},\ \bibinfo {pages} {2239} (\bibinfo {year} {1969})}\BibitemShut {NoStop}%
\bibitem [{\citenamefont {Callan}\ \emph {et~al.}(1969)\citenamefont {Callan}, \citenamefont {Coleman}, \citenamefont {Wess},\ and\ \citenamefont {Zumino}}]{Callan:1969sn}%
  \BibitemOpen
  \bibfield  {author} {\bibinfo {author} {\bibfnamefont {C.~G.}\ \bibnamefont {Callan}, \bibfnamefont {Jr.}}, \bibinfo {author} {\bibfnamefont {S.~R.}\ \bibnamefont {Coleman}}, \bibinfo {author} {\bibfnamefont {J.}~\bibnamefont {Wess}},\ and\ \bibinfo {author} {\bibfnamefont {B.}~\bibnamefont {Zumino}},\ }\href {https://doi.org/10.1103/PhysRev.177.2247} {\bibfield  {journal} {\bibinfo  {journal} {Phys. Rev.}\ }\textbf {\bibinfo {volume} {177}},\ \bibinfo {pages} {2247} (\bibinfo {year} {1969})}\BibitemShut {NoStop}%
\bibitem [{Note1()}]{Note1}%
  \BibitemOpen
  \bibinfo {note} {For the commuting broken generators, such terms cannot contribute to the Berry curvature.}\BibitemShut {Stop}%
\bibitem [{\citenamefont {Kobayashi}\ and\ \citenamefont {Nitta}(2014{\natexlab{a}})}]{Kobayashi:2014eqa}%
  \BibitemOpen
  \bibfield  {author} {\bibinfo {author} {\bibfnamefont {M.}~\bibnamefont {Kobayashi}}\ and\ \bibinfo {author} {\bibfnamefont {M.}~\bibnamefont {Nitta}},\ }\href {https://doi.org/10.1103/PhysRevD.90.025010} {\bibfield  {journal} {\bibinfo  {journal} {Phys. Rev. D}\ }\textbf {\bibinfo {volume} {90}},\ \bibinfo {pages} {025010} (\bibinfo {year} {2014}{\natexlab{a}})}\BibitemShut {NoStop}%
\bibitem [{\citenamefont {Kobayashi}\ and\ \citenamefont {Nitta}(2014{\natexlab{b}})}]{Kobayashi:2014xua}%
  \BibitemOpen
  \bibfield  {author} {\bibinfo {author} {\bibfnamefont {M.}~\bibnamefont {Kobayashi}}\ and\ \bibinfo {author} {\bibfnamefont {M.}~\bibnamefont {Nitta}},\ }\href {https://doi.org/10.1103/PhysRevLett.113.120403} {\bibfield  {journal} {\bibinfo  {journal} {Phys. Rev. Lett.}\ }\textbf {\bibinfo {volume} {113}},\ \bibinfo {pages} {120403} (\bibinfo {year} {2014}{\natexlab{b}})}\BibitemShut {NoStop}%
\bibitem [{Note2()}]{Note2}%
  \BibitemOpen
  \bibinfo {note} {See Ref.~\cite {GlodkowskiPhD} for details of the kernel construction. A closely related characterization was used in Ref.~\cite {Hayata:2013vfa} to elucidate the relation between broken symmetries and independent elastic variables.}\BibitemShut {Stop}%
\bibitem [{Note3()}]{Note3}%
  \BibitemOpen
  \bibinfo {note} {Loop corrections may generate a quantum scale anomaly in two spatial dimensions thereby lifting the flat radial direction.}\BibitemShut {Stop}%
\bibitem [{\citenamefont {Niederer}(1972)}]{Niederer:1972zz}%
  \BibitemOpen
  \bibfield  {author} {\bibinfo {author} {\bibfnamefont {U.}~\bibnamefont {Niederer}},\ }\href {https://doi.org/10.5169/seals-114417} {\bibfield  {journal} {\bibinfo  {journal} {Helv. Phys. Acta}\ }\textbf {\bibinfo {volume} {45}},\ \bibinfo {pages} {802} (\bibinfo {year} {1972})}\BibitemShut {NoStop}%
\bibitem [{\citenamefont {Hagen}(1972)}]{PhysRevD.5.377}%
  \BibitemOpen
  \bibfield  {author} {\bibinfo {author} {\bibfnamefont {C.~R.}\ \bibnamefont {Hagen}},\ }\href {https://doi.org/10.1103/PhysRevD.5.377} {\bibfield  {journal} {\bibinfo  {journal} {Phys. Rev. D}\ }\textbf {\bibinfo {volume} {5}},\ \bibinfo {pages} {377} (\bibinfo {year} {1972})}\BibitemShut {NoStop}%
\bibitem [{\citenamefont {Hayata}\ and\ \citenamefont {Hidaka}(2014)}]{Hayata:2013vfa}%
  \BibitemOpen
  \bibfield  {author} {\bibinfo {author} {\bibfnamefont {T.}~\bibnamefont {Hayata}}\ and\ \bibinfo {author} {\bibfnamefont {Y.}~\bibnamefont {Hidaka}},\ }\href {https://doi.org/10.1016/j.physletb.2014.06.039} {\bibfield  {journal} {\bibinfo  {journal} {Phys. Lett. B}\ }\textbf {\bibinfo {volume} {735}},\ \bibinfo {pages} {195} (\bibinfo {year} {2014})}\BibitemShut {NoStop}%
\end{thebibliography}%

\end{document}